\documentclass[aps,prl,12,groupedaddress,showpacs,letterpaper,twocolumn]{revtex4}

\usepackage{graphicx} 
\usepackage{amsmath}

\newcommand{\ket}[1]{|{#1}\rangle}

\begin{document}

\title{Deterministic entanglement of two neutral atoms via Rydberg blockade}

\author{X. L. Zhang, L. Isenhower, A. T. Gill, T. G. Walker, and M. Saffman}

\affiliation{Department of Physics, University of Wisconsin, 1150 University Avenue,
Madison, WI 53706 USA }
\begin{abstract}
We demonstrate  the first deterministic entanglement of two
individually addressed neutral atoms 
using a Rydberg blockade mediated controlled-NOT gate. 
Parity oscillation measurements reveal an  entanglement fidelity of $F=0.58\pm0.04$,  which is above
the entanglement threshold of $F=0.5$, without any correction for atom loss, and $F=0.71\pm0.05$ after correcting for background collisional losses. The fidelity results are shown to be in good agreement with a detailed error model. 
\end{abstract}


\maketitle
Entangled states are a crucial resource for quantum information processing (QIP) and quantum communication. 
Entangled states have been demonstrated with several different physical systems including trapped ions\cite{Sackett2000,Blatt2008}, superconductors\cite{Steffen2006}, photons\cite{Kwiat1995} and atomic systems\cite{Julsgaard2001,Mandel2003,Anderlini2007}. 
Recent experiments using Rydberg state mediated interactions of neutral atoms\cite{Jaksch2000} have demonstrated bipartite correlations  just below the threshold of $F=0.5$ for quantum entanglement\cite{Wilk2010,Isenhower2010}. 
Correction of the measured data for atom loss in those experiments revealed post-selected or {\it a posteriori} entanglement with fidelities of $F=0.58\pm.07$\cite{Isenhower2010} and $F=0.75\pm.07 $\cite{Wilk2010}. 
While {\it a posteriori} entanglement is useful for tests of Bell inequalities and studies of quantum non-locality, states with $F<0.5$ cannot be directly purified to cross the quantum boundary of $F>0.5.$ 

In this letter we report on Rydberg blockade experiments using improved experimental techniques that demonstrate deterministic preparation of entangled states with $F>0.5$ without any correction for atom loss during the
entire state preparation, gate operation, and measurement sequence. Applying a correction for losses due to background collisions that are independent of the gate operation we demonstrate entanglement fidelity of $F=0.71\pm.05.$ These results demonstrate 
unambiguously that Rydberg blockade can deterministically  generate entangled states which are a crucial resource for QIP.

It was first proposed in \cite{Jaksch2000} to use Rydberg
blockade interactions to implement a two-qubit entangling gate with neutral atoms.  The basic idea
behind  Rydberg blockade is straightforward: excitation of a control
atom to a Rydberg state prevents subsequent excitation of a target atom provided the dipole mediated Rydberg-Rydberg blockade shift $\sf B$ is 
large compared to the excitation Rabi frequency $\Omega$.
If the control atom excitation is state dependent we get a  
conditional phase shift on the target atom since excitation and de-excitation of the target
atom adds a  $\pi$ phase shift to the wavefunction.
This conditional phase gate $C_Z$ is converted 
into a  controlled-NOT gate ($H-C_{Z}$ CNOT) using
single qubit $\pi/2$ rotations (Hadamard gates) on the target atom before and
after the controlled phase operation. 

\begin{figure}[!t]
\includegraphics[width=.95\columnwidth]{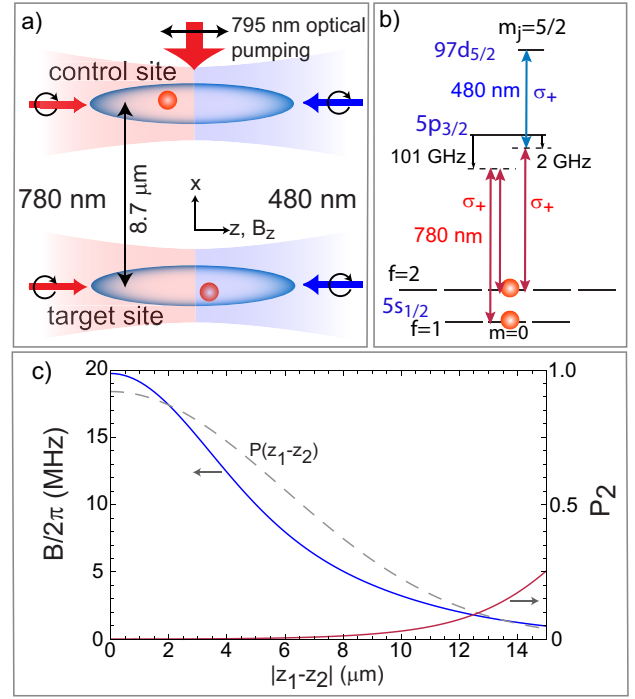}
\caption{(color online) a) Experimental geometry, b) relevant levels of $^{87}$Rb, and c) calculated blockade shift $\sf B$ and double excitation probability $P_2$. 
The relative probability distribution $P(|z_1-z_2|)$ assumes a trapping potential $U/k_B=4.5~\rm mK$,  waist of the 1064 nm trapping light of  $w=3.2~\mu\rm m$ and 
atom temperature $T=175~\mu\rm K$.
\label{fig.setup}}
\end{figure}

 The intrinsic error of the controlled phase operation can be estimated from\cite{Saffman2005a,Saffman2010}
\begin{equation}
E=\frac{7\pi}{4\Omega\tau}\left(1+\frac{\Omega^2}{\omega_{10}^2}+\frac{\Omega^2}{7{\sf B}^2} \right) + \frac{\Omega^2}{8{\sf B}^2}\left( 1+\frac{6{\sf B}^2}{\omega_{10}^2}\right).
\label{eq.error}
\end{equation}
In our experiments  $\tau\sim300~\mu\rm s$ is the radiative lifetime of the $97d_{5/2}$ Rydberg level, $\omega_{10}=2\pi\times 6.83~\rm GHz$ is the splitting of the $^{87}$Rb qubit basis states, and $\Omega=2\pi\times  
0.81 ~\rm MHz$. In the experimental geometry shown in Fig. \ref{fig.setup} a range of two-atom separations, and hence blockade shifts, occur.
 The blockade shift curve shown in Fig. \ref{fig.setup}c was calculated from the theory of Ref. \cite{Walker2008} using 
a trap separation of $x=8.7~\mu\rm m$ and a bias magnetic field of 
$B_z=0.37~\rm mT$ applied along the $\hat z$ axis. Averaging  Eq. (\ref{eq.error}) over the probability distribution  $P(|z_1-z_2|)$ gives an expected error of $E=6.5\times 10^{-3}$.
The corresponding averaged blockade shift from Eq. (\ref{eq.error}) is ${\sf B}=2\pi\times 5.3~\rm MHz$.  
The spatial distribution $P(|z_1-z_2|)$ is dependent on the trapped atom temperature which is estimated to be  
$175~\mu\rm K$ based on compatibility of three different diagnostics: release and recapture measurements\cite{Reymond2003}, comparison of 
the atomic spatial distribution found from averaged camera images  to that expected given our trapping beam waist and power\cite{Urban2009}, and measurement of the radial vibrational frequency from parametric heating data. Uncertainties in the beam waist and trapping power lead to a  $\sim\pm20\%$ uncertainty in the atom temperature. This uncertainty does not significantly affect the estimated intrinsic gate error. 

This calculation shows that the intrinsic error of the Rydberg blockade gate is less than 1\% with available experimental parameters. 
As we show below the observed gate fidelity errors are a factor of 10 
or more higher than our theoretical limit, the reason being that 
Eq. (\ref{eq.error}) does not account for all experimental effects and technical imperfections.
Of particular importance are Doppler effects which both degrade the Rydberg excitation fidelity and add
a stochastic phase which limits the entanglement fidelity. 
    In what follows we  present experimental results showing improved CNOT truth table fidelity and deterministic two-atom entanglement. We then identify the dominant error sources leading to loss of fidelity, and suggest ways in which the theoretical error limit could be reached in future work.

The experimental apparatus and procedures have been described in detail
in our recent publications \cite{Urban2009,Isenhower2010}. Here we describe the  procedures briefly and highlight the changes relative to our previous work. Single atoms with temperatures of $T\sim 250~\mu\rm K$ are loaded from a magneto-optical trap into 
optical traps created by focusing 1064 nm light to waists of $w=3.2~\mu\rm m$. Using a diffraction grating we create a linear array of 5 trapping sites which can be individually controlled\cite{Knoernschild2010}. In the present work we use two of the central sites with a separation of $x=8.7~\mu\rm m$. This is slightly smaller than in our previous CNOT experiment ($x=10~\mu\rm m$)  which gives about 
$2.3$ times more blockade shift in the van der Waals limit. 
A bias magnetic field is applied along $z$, which defines the quantization
axis for optical pumping (using $B_z=0.15~\rm mT$) and lifts the degeneracy
of the Rydberg state Zeeman sublevels (using $B_z=0.37~\rm mT$) during 
gate operation. 

\begin{figure}[!t]
\includegraphics[width=1\columnwidth]{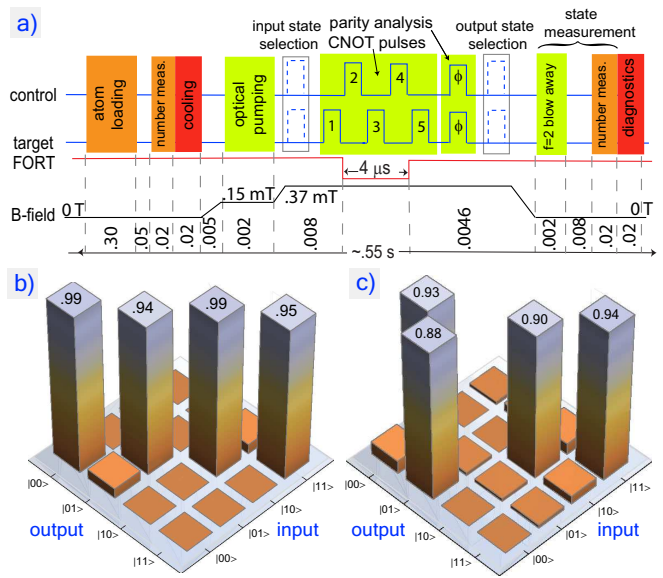}
\vspace{-1.cm}
\caption{(color online) a) Experimental sequence, pulses 1 and 5 are ground
Rabi $\pi/2$ pulses; pulses 2 and 4 are Rydberg $\pi$ pulses; pulse
3 is a Rydberg $2\pi$ pulse. Measured probabilities for state preparation b) and  $H-C_{Z}$ CNOT c).
The reported matrices are based
on an average of about 100 data points for each matrix element. 
\label{fig.cnot}}
\end{figure}

The experimental sequence is shown in Fig. \ref{fig.cnot}a. After loading into the optical traps an atom number measurement is made to verify the presence of two atoms, followed by a $20 ~\rm ms$ cooling phase (light detuned by $-3\gamma$ from the cycling transition). The cooling phase gives $T\sim 175\pm 35 ~\mu\rm K $ atoms which are then 
 optically pumped into 
$|f=2,m_{f}=0\rangle$ using $\pi$
polarized light propagating along $-x$ tuned to the $|5s_{1/2},f=2\rangle\rightarrow|5p_{1/2},f'=2\rangle$ D1 transition
at $795~\rm nm$ and $780~\rm nm$ light tuned 
to the $|5s_{1/2},f=1\rangle\rightarrow|5p_{3/2},f'=2\rangle$
D2 transition. The 
$5p_{1/2}$
level has a larger excited state hyperfine splitting than $5p_{3/2}$ and we thereby obtain an improved optical pumping efficiency compared to \cite{Isenhower2010} of about $0.99.$
Readout of the qubit state of the atoms is performed by using light forces to remove one of the hyperfine states\cite{Isenhower2010} and then collecting
resonance fluorescence on a cooled EMCCD camera from a region of interest
with predetermined thresholds indicating the presence or absence of
a single atom. Typical integration times were approximately $10~\rm ms$. 

\begin{figure}
\includegraphics[width=1\columnwidth]{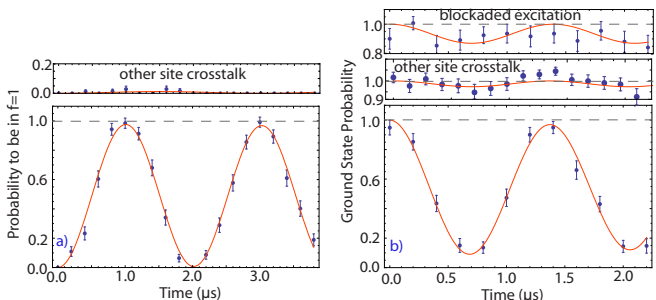}
\vspace{-.8cm}
\caption{(color online) a) Ground Rabi flopping on targeted site with neighboring site crosstalk in upper panel and b) Rydberg Rabi flopping on targeted site, with neighboring site crosstalk and  flopping blocked by prior excitation of the neighboring site  in upper panels. 
 The flopping 
curves are based on an average of about 100 measurements for each point.\label{fig.flopping}}
\end{figure}

For the qubit basis we use the ground hyperfine states $\ket{0}\equiv\ket{f=1,m_f=0}, \ket{1}\equiv\ket{f=2,m_f=0}.$
Single qubit rotations  $\ket{0}\leftrightarrow\ket{1}$  use
 two-photon stimulated Raman transitions driven by focusing a $\sigma_{+}$ polarized $780~\rm nm$
laser with frequency components separated by $\omega_{10}$ and
detuned by $\Delta/2\pi=-101~{\rm GHz}$ to the red of the D2 transition\cite{Yavuz2006}. Typical total power in the two
Raman sidebands is $\sim90{\rm ~}\mu{\rm W}$ and we achieve $\pi$
pulse times of $\sim900~\rm ns$ as shown in Fig. \ref{fig.flopping}.   with peak-peak amplitude of better than 0.98 
after correction for background atom loss of $10\%$. The single atom  loss probability during the gap of $0.11~\rm s$ between state preparation and output state measurement   has been reduced from $15\%$ in earlier work to typically $10\%$ (see Fig. \ref{fig.cnot}a). This was achieved using faster mechanical shutters for laser beams and faster magnetic field switching electronics. 

Coherent Rydberg excitation between   $|1\rangle$ and $|r\rangle=\ket{97d_{5/2},m_j=5/2}$ uses a two photon transition with
$\sigma_{+}$ polarized 780 and $480~\rm nm$ beams. The $780~\rm nm$
beam is tuned by about $\Delta_{f=2}=-2\pi\times 2~{\rm GHz}$ to the red of the D2 line. Typical beam powers are $2.4~{\mu\rm W}(13~{\rm mW})$ with $x-y$ averaged beam waists of $7.7(4.5)~\mu\rm m$ at $780(480)~\rm nm$
 giving Rydberg $\pi$ pulse
times of $\sim620~\rm ns$ with peak-peak amplitude from a least squares fit  of $0.92$ after correction for background atom loss of $10\%$, as shown in Fig.\ref{fig.flopping}b. Improved long term stability of the pulse areas for ground state and Rydberg operations has been established using a system that monitors the powers transmitted through single mode fibers to the experimental chamber ($20~\rm ms$ diagnostic slice at the end of each sequence in Fig. \ref{fig.cnot}a), and corrects  
the incident beam powers accordingly\cite{Gill2010}.

Using the above methods we have generated the CNOT probability truth table shown in Fig. \ref{fig.cnot}. As input to the CNOT sequence we apply the four two-qubit computational states which can be prepared with an average fidelity of $0.97$, as shown in Fig. \ref{fig.cnot}b). 
The measured  CNOT probability truth table  shown in Fig. \ref{fig.cnot}c has the values
\begin{equation}
|U_{{\rm CNOT}}|\hspace{-.15cm}=\hspace{-.15cm}\left(\begin{array}{cccc}
0.08&0.93\pm.06 & 0  & 0\\
0.88\pm.06 & 0.02 & 0.02 & 0.02\\
0  &0 & 0.90\pm.07& 0.05 \\
0.02 & 0.05 & 0.07&0.94\pm.06\end{array}\right).\nonumber
\end{equation}
The
CNOT table was obtained using ground state $\pi/2$ pulses that were $\pi$
out of phase which puts the large off-diagonal values in the upper left quadrant. 
Correcting for background atom loss the fidelity is $F=\frac{1}{4}{\rm Tr}[|U_{{\rm ideal}}^{T}|U_{{\rm CNOT}}]=0.91\pm .06$.  As can be seen from Table \ref{tab.results} the background and trace loss corrected CNOT fidelity of $0.92$ agrees to within a few percent with the expected error of $1-0.06=0.94.$ The error budget for the CNOT gate stems from several sources with the dominant errors due to spontaneous emission 
 and Doppler broadening. Both of these errors could be reduced with 
a higher power 480 nm laser which would allow for faster Rydberg excitation together with larger detuning from the intermediate $5p_{3/2}$ level. Since $\Omega\sim P_{\rm se} I_{480}$ where $P_{\rm se}$ is the probability of spontaneous emission during excitation and $I_{480}$ is the intensity of the 480 nm laser\cite{Saffman2010} simultaneously decreasing the spontaneous emission and the Doppler broadening errors each by a factor of ten could be achieved with an increase of about 30 times in laser intensity. 

\begin{table}[!t]
\caption{Error sources for the CNOT truth table and entanglement results. 
Spontaneous emission refers to scattering from $5p_{3/2}$ during Rydberg excitation and Doppler broadening causes imperfect Rydberg excitation.
\label{tab.results}}
\begin{tabular}{lcc}
\hline 
{\bf error sources (two qubits)} & Ref. \cite{Isenhower2010}&this work\tabularnewline
\hline\hline
optical pumping  & $0.1$ & $0.02$\tabularnewline
\hline
{\footnotesize atom losses before CNOT pulses}  & $0.09$ & $0.02$\tabularnewline
\hline
{\footnotesize blockade error at $175~\mu\rm K$} & $0.01$ & $0.01$\tabularnewline
\hline
{\footnotesize spontaneous emission}  & $0.04$ & $0.04$\tabularnewline
\hline
{\footnotesize Doppler broadening}  & $0.04$ & $0.04$\tabularnewline
\hline
{\footnotesize Total CNOT error (added in quadrature)}  & $\sim0.15$ & $\sim0.06$\tabularnewline
\hline 
{\bf measured results} & &\tabularnewline
\hline\hline
{\footnotesize background loss (two atoms)} & $0.28$ & $0.19$\tabularnewline
\hline
gate trace loss $(1-{\rm Tr}[\rho])$ & $0.17$ & $\sim0.01$\tabularnewline
\hline\hline
CNOT fidelity raw & $0.52$ & $0.74$\tabularnewline
\hline
{\footnotesize CNOT  background loss corrected} & $0.72$ & $0.91$\tabularnewline
\hline
{\footnotesize  CNOT background \& trace corrected} & $0.86$ & $0.92$\tabularnewline
\hline
\hline
{\footnotesize entanglement fidelity raw} & $0.34$ & $0.58$\tabularnewline
\hline
{\footnotesize entanglement background loss corrected} & $0.48$ & $0.71$\tabularnewline
\hline
{\footnotesize entanglement background \& trace corrected} & $0.58$ & $0.71$\tabularnewline
\hline
\end{tabular}
\end{table}

\begin{figure}[!t]
\includegraphics[width=1\columnwidth]{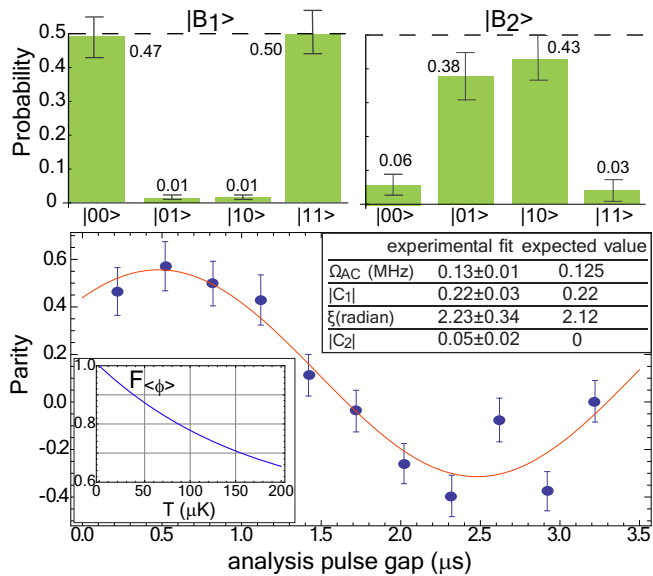}
\vspace{-.6cm}
\caption{(color online) Measured probabilities for preparation of Bell states $|B_{1}\rangle,|B_{2}\rangle$ and parity oscillation signal  obtained from
$|B_{1}\rangle$. 
The upper inset gives parameters found from fitting the observed parity signal together with expected values derived from measured experimental parameters. The lower inset shows the stochastic phase limited fidelity vs. temperature  (see text for details). 
\label{fig.parity}}
\end{figure}

To  create entangled states 
we used $\pi/2$ pulses on the control atom to prepare the input states
$|ct\rangle=\frac{1}{\sqrt{2}}(|0\rangle+i|1\rangle)|1\rangle$ and
$|ct\rangle=\frac{1}{\sqrt{2}}(|0\rangle+i|1\rangle)|0\rangle$. Applying
the CNOT to these states creates two of the Bell states $|B_{1}\rangle=\frac{1}{\sqrt{2}}(|00\rangle+|11\rangle)$
and $|B_{2}\rangle=\frac{1}{\sqrt{2}}(|01\rangle+|10\rangle)$. The
measured probabilities for these output states are shown in Fig. \ref{fig.parity}.
In order to verify entanglement of $|B_1\rangle$ we measured the
parity signal\cite{Turchette1998} $P=P_{00}+P_{11}-P_{01}-P_{10}$ after applying  $\pi/2$ analysis pulses to both atoms between the last CNOT pulse
and the state selection pulse. The analysis pulses had  a variable 
phase $\phi=\Omega_{AC}t$, 
where $\Omega_{AC}=2\pi\times 0.125~\rm MHz$ is due to AC Stark shifts from the ground state Raman beams, and $t$ is the analysis pulse gap in Fig. \ref{fig.parity}.

A short calculation shows that the parity signal should vary as $P=2{\rm Re}(C_{2})-2|C_{1}|\cos(2\phi+\xi)$
where $C_{2}$ is the
coherence between states $|01\rangle$ and $|10\rangle$, and $C_{1}=|C_{1}|e^{\imath\xi}$
is the coherence between states $|00\rangle$ and $|11\rangle$. 
The parameters extracted from a least squares curve fit are 
given in Fig. \ref{fig.parity}.
The parity signal  oscillates at a frequency 2$\Omega_{AC}=2\pi\times 0.26~\rm MHz$
which is within $5\%$ of the expected value. The phase $\xi$ of $C_{1}$ can be
explained by the differential AC Stark shift induced during Rydberg excitation.
Accounting for all three Rydberg pulses gives a differential phase shift between
$|00\rangle$ and $|11\rangle$ of 
$\xi=-2\pi \frac{\Omega_{780}}{\Omega_{480}}\frac{\omega_{10}}{\Delta_{f=1}}=2.12+4\pi$
which agrees with the experimental result to within  $10\%$. Here  $\Omega_{780}=2\pi\times 118{\rm ~}{\rm MHz}$, $\Omega_{480}=2\pi\times 39{\rm ~}{\rm MHz}$
are calculated one-photon  Rydberg Rabi frequencies,  and $\Delta_{f=1}=\Delta_{f=2}-\omega_{10}.$

 The fidelity of entanglement of the Bell state $|B_{1}\rangle$
can be quantified by $F=\frac{1}{2}(P_{00}+P_{11})+|C_{1}|$ \cite{Sackett2000}.
A sufficient, but not necessary, condition for entanglement is  $0.5<F\le1$. The population data in Fig. \ref{fig.parity} together with the measured coherence 
yield $F=0.58\pm0.04$ without applying any corrections for atom loss or trace loss. This is the first demonstration of deterministic entanglement using Rydberg blockade. The corrected fidelities shown in Table \ref{tab.results} reach $F=0.71\pm.05$ which is above the entanglement threshold by more than $4\sigma$.  
The entanglement fidelity of $|B_2\rangle$ was not measured, but we would expect it to be somewhat lower based on the population data in Fig. \ref{fig.parity}.

It is apparent that the fidelity of the entangled state is lower than that of the CNOT truth table, and lower than that expected from the error budget in Table \ref{tab.results}. This is due to the fact that the entanglement fidelity  is sensitive to errors that do not affect  the CNOT probabilities. As was pointed out in \cite{Wilk2010} the motion of 
 Rydberg excited atoms between  excitation and deexcitation 
pulses leads to a stochastic phase $\varphi$ that degrades the entanglement fidelity. In our implementation  $\varphi=({\bf k}\cdot{\bf v}) t_{24}$ where $|{\bf k}|=2\pi/\lambda_{480}-2\pi/\lambda_{780}$, 
$\bf v$ is the atomic velocity, and $t_{24}=2.2~\mu\rm s$ is the gap time between pulses 2 and 4 in Fig. \ref{fig.cnot}. Assuming  a temperature of $150~\mu\rm K$, which is close to the lower end of our experimental 
range $(140~\mu{\rm K}< T < 210 ~\mu{\rm K})$, and averaging over the thermal velocity 
distribution   gives $\left\langle e^{\imath\varphi}\right\rangle =0.41 $ which implies a maximum fidelity of 
$F_{\rm <\varphi>}=0.71.$ Adding to this the total CNOT error from Table \ref{tab.results}  gives $F_{\rm max}\sim 0.65$ which is somewhat lower than our experimental result. As is seen in the inset of Fig. \ref{fig.parity} the fidelity limit set by $F_{\rm <\varphi>}$
depends strongly on temperature.  
Calculations show  that reducing the atom temperature to $50~\mu\rm K$ and the gap time to $t_{24}=1.5~\mu\rm s$ would result in an
entanglement fidelity above $0.9.$

In conclusion we have deterministically generated  entangled states
of two neutral atoms 
using a Rydberg blockade mediated CNOT gate. Entanglement fidelity 
of $F=0.58$ is obtained without any corrections for atom loss, and $F=0.71$ after accounting for losses due to collisions with background atoms. The observed fidelities are shown to be in reasonable agreement with an error model that takes into account experimental imperfections as well as the finite temperature of the atoms. 
Our results suggest that it should be possible to reach fidelities above  $0.9$ by increasing the power of the Rydberg excitation laser and better cooling of the atoms.

This work was supported by NSF grant PHY-0653408 and ARO/IARPA under
contract W911NF-05-1-0492.

%
%


\begin{thebibliography}{19}
\expandafter\ifx\csname natexlab\endcsname\relax\def\natexlab#1{#1}\fi
\expandafter\ifx\csname bibnamefont\endcsname\relax
  \def\bibnamefont#1{#1}\fi
\expandafter\ifx\csname bibfnamefont\endcsname\relax
  \def\bibfnamefont#1{#1}\fi
\expandafter\ifx\csname citenamefont\endcsname\relax
  \def\citenamefont#1{#1}\fi
\expandafter\ifx\csname url\endcsname\relax
  \def\url#1{\texttt{#1}}\fi
\expandafter\ifx\csname urlprefix\endcsname\relax\def\urlprefix{URL }\fi
\providecommand{\bibinfo}[2]{#2}
\providecommand{\eprint}[2][]{\url{#2}}

\bibitem[{\citenamefont{Sackett et~al.}(2000)\citenamefont{Sackett, Kielpinski,
  King, Langer, Meyer, Myatt, Rowe, Turchette, Itano, Wineland
  et~al.}}]{Sackett2000}
\bibinfo{author}{\bibfnamefont{C.~A.} \bibnamefont{Sackett}},
  \bibnamefont{et~al.}, \bibinfo{journal}{Nature (London)}
  \textbf{\bibinfo{volume}{404}}, \bibinfo{pages}{256} (\bibinfo{year}{2000}).

\bibitem[{\citenamefont{Blatt and Wineland}(2008)}]{Blatt2008}
\bibinfo{author}{\bibfnamefont{R.}~\bibnamefont{Blatt}} \bibnamefont{and}
  \bibinfo{author}{\bibfnamefont{D.}~\bibnamefont{Wineland}},
  \bibinfo{journal}{Nature (London)} \textbf{\bibinfo{volume}{453}},
  \bibinfo{pages}{1008} (\bibinfo{year}{2008}).

\bibitem[{\citenamefont{Steffen et~al.}(2006)\citenamefont{Steffen, Ansmann,
  Bialczak, Katz, Lucero, McDermott, Neeley, Weig, Cleland, and
  Martinis}}]{Steffen2006}
\bibinfo{author}{\bibfnamefont{M.}~\bibnamefont{Steffen}},  \bibnamefont{et~al.}, \bibinfo{journal}{Science}
  \textbf{\bibinfo{volume}{313}}, \bibinfo{pages}{1423} (\bibinfo{year}{2006}).

\bibitem[{\citenamefont{Kwiat et~al.}(1995)\citenamefont{Kwiat, Mattle,
  Weinfurter, Zeilinger, Sergienko, and Shih}}]{Kwiat1995}
\bibinfo{author}{\bibfnamefont{P.~G.} \bibnamefont{Kwiat}},  \bibnamefont{et~al.},
  \bibinfo{journal}{Phys. Rev. Lett.} \textbf{\bibinfo{volume}{75}},
  \bibinfo{pages}{4337} (\bibinfo{year}{1995}).

\bibitem[{\citenamefont{Julsgaard et~al.}(2001)\citenamefont{Julsgaard,
  Kozhekin, and Polzik}}]{Julsgaard2001}
\bibinfo{author}{\bibfnamefont{B.}~\bibnamefont{Julsgaard}},
  \bibinfo{author}{\bibfnamefont{A.}~\bibnamefont{Kozhekin}}, \bibnamefont{and}
  \bibinfo{author}{\bibfnamefont{E.~S.}~\bibnamefont{Polzik}},
  \bibinfo{journal}{Nature (London)} \textbf{\bibinfo{volume}{413}},
  \bibinfo{pages}{400} (\bibinfo{year}{2001}).

\bibitem[{\citenamefont{Mandel et~al.}(2003)\citenamefont{Mandel, Greiner,
  Widera, Rom, H\"ansch, and Bloch}}]{Mandel2003}
\bibinfo{author}{\bibfnamefont{O.}~\bibnamefont{Mandel}},  \bibnamefont{et~al.},
  \bibinfo{journal}{Nature (London)} \textbf{\bibinfo{volume}{425}},
  \bibinfo{pages}{937} (\bibinfo{year}{2003}).

\bibitem[{\citenamefont{Anderlini et~al.}(2007)\citenamefont{Anderlini, Lee,
  Brown, Sebby-Strabley, Phillips, and Porto}}]{Anderlini2007}
\bibinfo{author}{\bibfnamefont{M.}~\bibnamefont{Anderlini}},  \bibnamefont{et~al.},
  \bibinfo{journal}{Nature (London)} \textbf{\bibinfo{volume}{448}},
  \bibinfo{pages}{452} (\bibinfo{year}{2007}).

\bibitem[{\citenamefont{Jaksch et~al.}(2000)\citenamefont{Jaksch, Cirac,
  Zoller, Rolston, C\^ot\'e, and Lukin}}]{Jaksch2000}
\bibinfo{author}{\bibfnamefont{D.}~\bibnamefont{Jaksch}},  \bibnamefont{et~al.},
  \bibinfo{journal}{Phys. Rev. Lett.} \textbf{\bibinfo{volume}{85}},
  \bibinfo{pages}{2208} (\bibinfo{year}{2000}).

\bibitem[{\citenamefont{Wilk et~al.}(2010)\citenamefont{Wilk, Ga\"etan,
  Evellin, Wolters, Miroshnychenko, Grangier, and Browaeys}}]{Wilk2010}
\bibinfo{author}{\bibfnamefont{T.}~\bibnamefont{Wilk}},  \bibnamefont{et~al.},
  \bibinfo{journal}{Phys. Rev. Lett.} \textbf{\bibinfo{volume}{104}},
  \bibinfo{pages}{010502} (\bibinfo{year}{2010}).

\bibitem[{\citenamefont{Isenhower et~al.}(2010)\citenamefont{Isenhower, Urban,
  Zhang, Gill, Henage, Johnson, Walker, and Saffman}}]{Isenhower2010}
\bibinfo{author}{\bibfnamefont{L.}~\bibnamefont{Isenhower}},  \bibnamefont{et~al.},
  \bibinfo{journal}{Phys. Rev. Lett.} \textbf{\bibinfo{volume}{104}},
  \bibinfo{pages}{010503} (\bibinfo{year}{2010}).

\bibitem[{\citenamefont{Saffman and Walker}(2005)}]{Saffman2005a}
\bibinfo{author}{\bibfnamefont{M.}~\bibnamefont{Saffman}} \bibnamefont{and}
  \bibinfo{author}{\bibfnamefont{T.~G.} \bibnamefont{Walker}},
  \bibinfo{journal}{Phys. Rev. A} \textbf{\bibinfo{volume}{72}},
  \bibinfo{pages}{022347} (\bibinfo{year}{2005}).

\bibitem[{\citenamefont{Saffman et~al.}(2009)\citenamefont{Saffman, Walker, and
  M\o{}lmer}}]{Saffman2010}
\bibinfo{author}{\bibfnamefont{M.}~\bibnamefont{Saffman}},
  \bibinfo{author}{\bibfnamefont{T.~G.} \bibnamefont{Walker}},
  \bibnamefont{and}
  \bibinfo{author}{\bibfnamefont{K.}~\bibnamefont{M\o{}lmer}},
  \bibinfo{journal}{arXiv:0909.4777}  (\bibinfo{year}{2009}).

\bibitem[{\citenamefont{Walker and Saffman}(2008)}]{Walker2008}
\bibinfo{author}{\bibfnamefont{T.~G.} \bibnamefont{Walker}} \bibnamefont{and}
  \bibinfo{author}{\bibfnamefont{M.}~\bibnamefont{Saffman}},
  \bibinfo{journal}{Phys. Rev. A} \textbf{\bibinfo{volume}{77}},
  \bibinfo{pages}{032723} (\bibinfo{year}{2008}).

\bibitem[{\citenamefont{Reymond et~al.}(2003)\citenamefont{Reymond, Schlosser,
  Protsenko, and Grangier}}]{Reymond2003}
\bibinfo{author}{\bibfnamefont{G.}~\bibnamefont{Reymond}},
  \bibinfo{author}{\bibfnamefont{N.}~\bibnamefont{Schlosser}},
  \bibinfo{author}{\bibfnamefont{I.}~\bibnamefont{Protsenko}},
  \bibnamefont{and} \bibinfo{author}{\bibfnamefont{P.}~\bibnamefont{Grangier}},
  \bibinfo{journal}{Phil. Trans. R. Soc. Lond. A}
  \textbf{\bibinfo{volume}{361}}, \bibinfo{pages}{1527} (\bibinfo{year}{2003}).

\bibitem[{\citenamefont{Urban et~al.}(2009)\citenamefont{Urban, Johnson,
  Henage, Isenhower, Yavuz, Walker, and Saffman}}]{Urban2009}
\bibinfo{author}{\bibfnamefont{E.}~\bibnamefont{Urban}},  \bibnamefont{et~al.},
  \bibinfo{journal}{Nature Phys.} \textbf{\bibinfo{volume}{5}},
  \bibinfo{pages}{110} (\bibinfo{year}{2009}).

\bibitem[{\citenamefont{Knoernschild et~al.}(2010)\citenamefont{Knoernschild,
  Zhang, Isenhower, Gill, Lu, Saffman, and Kim}}]{Knoernschild2010}
\bibinfo{author}{\bibfnamefont{C.}~\bibnamefont{Knoernschild}},  \bibnamefont{et~al.},
  \bibinfo{journal}{arXiv:1006.2757}  (\bibinfo{year}{2010}).

\bibitem[{\citenamefont{Yavuz et~al.}(2006)\citenamefont{Yavuz, Kulatunga,
  Urban, Johnson, Proite, Henage, Walker, and Saffman}}]{Yavuz2006}
\bibinfo{author}{\bibfnamefont{D.~D.} \bibnamefont{Yavuz}},  \bibnamefont{et~al.},
  \bibinfo{journal}{Phys. Rev. Lett.} \textbf{\bibinfo{volume}{96}},
  \bibinfo{pages}{063001} (\bibinfo{year}{2006}).

\bibitem[{\citenamefont{Gill}(2010)}]{Gill2010}
\bibinfo{author}{\bibfnamefont{A.~T.} \bibnamefont{Gill}},
  \bibinfo{journal}{unpublished}  (\bibinfo{year}{2010}).

\bibitem[{\citenamefont{Turchette et~al.}(1998)\citenamefont{Turchette, Wood,
  King, Myatt, Leibfried, Itano, Monroe, and Wineland}}]{Turchette1998}
\bibinfo{author}{\bibfnamefont{Q.~A.} \bibnamefont{Turchette}},  \bibnamefont{et~al.},
  \bibinfo{journal}{Phys. Rev. Lett.} \textbf{\bibinfo{volume}{81}},
  \bibinfo{pages}{3631} (\bibinfo{year}{1998}).

\end{thebibliography}

\end{document}